\documentstyle[prc,aps,amsmath,preprint]{revtex}

\begin{document}
\draft
\title{Relativistic Hartree-Bogoliubov Approach for Nuclear Matter
with Non-Linear Coupling Terms}
\author{S.~Sugimoto$^{a,b,}$\footnote{e-mail: satoru@postman.riken.go.jp}, 
K.~Sumiyoshi$^{c,}$\footnote{e-mail: sumi@la.numazu-ct.ac.jp},
and 
H.~Toki$^{a,b,}$\footnote{e-mail: toki@rcnp.osaka-u.ac.jp}
}
\address{
      $^{a}$The Institute of Physical and Chemical Research (RIKEN), \\
            Hirosawa, Wako, Saitama 351-0198, Japan \\
      $^{b}$Research Center for Nuclear Physics(RCNP), Osaka University, \\
            Mihogaoka, Ibaraki, Osaka 567-0047, Japan \\
      $^{c}$Numazu College of Technology,\\
      Ooka, Numazu, Shizuoka 410-8501, Japan
}
\date{\today}
\maketitle
\begin{abstract}
We investigate the pairing property of nuclear matter with
Relativistic Hartree-Bogoliubov(RHB)
approach.
Recently, the RHB approach has been widely applied to nuclear matter
and finite nuclei.
We have extended the RHB approach to be able to include non-linear
coupling terms of mesons.
In this paper we apply it to nuclear matter and observe the effect
of non-linear terms on pairing gaps.
\end{abstract}
\pacs{21.60.-n, 21.65.+f}
\section{Introduction}
The development of secondary beam to produce
unstable nuclei changed the common sense of nuclear physics, which had been
accumulated over the years from the study of stable nuclei.
The discovery of neutron halo and skin is one of the most famous ones.
These observations changed the formula of the nuclear radius being
proportional to A$^{1/3}$.
New facilities are now planned and being constructed to get more
unstable nuclei including
those near the drip line.
These new facilities would definitely bring us with more information
and new phenomena for those unstable nuclei.

In the theory side, various attempts were made to understand diverse
phenomena with the unified frameworks.
One of the powerful framework is the relativistic mean field theory(RMF)
\cite{SW86,Re89,GRT90,Ri96}.
It is based on the relativistic many-body theory and is applied to many
observables with great success.

In the RMF theory, it is very important to include the
pairing interaction on the same footing as the particle-hole correlations.
Usually, it is implemented by the so-called
BCS-type theory as in other mean field theory.
In the BCS-type theory the pairing interaction is implemented by
hand outside of the relativistic theory.
However, we would like that the pairing interaction is
described in the relativistic many-body theoretical manner within the
RMF theory.
One possibility toward this attempt is the relativistic
Hartree-Fock-Bogoliubov theory(RHFB), which was formulated by Kucharek et al
\cite{KR91}. Recently, the Hartree-Fock-Bogoliubov(HFB) theory is
widely used  
in both the non-relativistic and relativistic mean field calculations.
It can treat the usual mean field and the pairing field simultaneously.
The HFB theory is applied to nuclear matter and finite nuclei, 
relativistically \cite{KR91,GCF96,MFD97,Ma98,MT99,MR96,LVPR98,LVR98} and 
non-relativistically \cite{KPS89,DFT84,DNWBCD96} with many interesting results.

However, until now the effect of the non-linear self-coupling terms of 
mesons are not considered in the pairing channel in the RHB theory.
It is known that 
those non-linear terms are needed to reproduce the properties of nuclear matter and
finite nuclei in the RMF theory \cite{BB77,Bo91,ST94}.
We can imagine that those terms have some effect on the pairing correlations.
Therefore,
in this paper we would like to  extend  the
relativistic Hartree-Fock-Bogoliubov
theory to include the effect of the non-linear self coupling terms of
mesons and apply it to nuclear matter to see the effect
on the pairing gap $\Delta$.

In Section~\ref{sec: formalism}, we develop the formalism of the RHFB
theory using the non-linear relativistic Lagrangian under the harmonic
approximation to handle the non-linear terms.
In Section~\ref{sec: result},
this equation is solved with the parameter sets of TM1\cite{ST94} and
NL1 \cite{RRM86}.
Since these parameter sets were obtained in the RMF formulation, the
Fock term contribution is dropped in this paper.
We study the form factor dependence on the pairing properties of the
nuclear matter.
Section~\ref{sec: conc} is devoted to the summary of this paper.
\section{Formalism}\label{sec: formalism}
In the RMF theory we use the Lagrangian density
\cite{SW86,Re89,GRT90,Ri96,ST94},
\begin{align}
{\mathcal{L}} \left(x\right)
&= \bar{\psi} \left(x\right) \left[ i \gamma_\mu \partial^\mu - M
- g_\sigma \sigma - g_\omega \gamma_\mu \omega^\mu
-g_\rho \gamma_\mu \tau^a \rho^{\mu a} - e \gamma_\mu \frac{1-\tau_3}{2} A^\mu
\right] \psi \left(x\right)
\notag \\
&+ \frac{1}{2} \partial_\mu \sigma \left(x\right)
\partial^\mu \sigma \left(x\right) - U\left(\sigma\right)
- \frac{1}{4} V_{\mu\nu} V^{\mu\nu}
+ \frac{1}{2} m_\omega^2 \omega_\mu \omega^\mu
+ \frac{1}{4} c_3 \left(\omega_\mu \omega^\mu \right)^2
\notag \\
&-\frac{1}{4} \vec{R}_{\mu\nu} \cdot \vec{R}^{\mu\nu}
+ \frac{1}{2} m_\rho^2 \vec{\rho}_\mu \cdot \vec{\rho}\ ^\mu
-\frac{1}{4} F_{\mu\nu} F^{\mu\nu}.
\label{eq: Lag}
\end{align}
In this Lagrangian $\psi$ is a Dirac spinor of a nucleon.
We introduce three types of meson fields $\sigma$, $\omega$ and $\rho$.
$\sigma$ is the isoscalar-scalar field, which mediates medium range
attraction between two nucleons.
$\omega$ is the isoscalar-vector field, which produces short range
repulsion.
$\rho$ is the isovector-vector field, which gives isovector correlation
among nucleons in
nuclei.
$A$ is the photon field, which produces the electromagnetic field.
Here, $U(\sigma)$ is self coupling terms of the $\sigma$ meson,
\begin{align}
U \left(\sigma\right) & = \frac{1}{2} m_\sigma^2 \sigma^2
+ \frac{1}{3} g_2 \sigma^3
+ \frac{1}{4} g_3 \sigma^4.
\end{align}
Boguta et al. investigated it extensively \cite{BB77} and found that it
is needed to 
reproduce the nuclear matter and finite nucleus properties.
We also include a non-linear coupling term of $\omega$ meson \cite{Bo91,ST94}, which was
investigated by Sugahara et al.,  motivated by the recent success of the
relativistic Br\"{u}ckner-Hartree-Fock(RBHF) theory\cite{BM90}.
They found that it is
needed to reproduce the RBHF result for the nuclear matter
quantitatively up to the high density region.
From this Lagrangian, we can derive a Dirac equation for nucleons and
Klein-Gordon equations for bosons with the Euler-Lagrange equation.
In the RMF theory we replace the meson fields with their classical mean
values,
\begin{align}
 \sigma \rightarrow \left< \sigma \right> \quad \text{etc}.
\end{align}
However, in this approximation the terms which lead to the pairing
interaction do not appear. These terms come from the terms like $\psi^\dagger\psi^\dagger\psi\psi$.
To obtain these terms we need to take the higher order terms and
fluctuation terms of the meson fields \cite{KR91}.
Here we explain our method which is somewhat different from Kucharek et
al.
For brevity, we write down  only the $\sigma$ meson terms. 
The extension to include  other mesons
is straightforward.
We write the quantized field as the sum of its classical field and its
fluctuation,
\begin{align}
 \sigma = \bar{\sigma} + \sigma '.
\label{eq: divide}
\end{align}
Here, the classical mean value $\bar{\sigma}$ is obtained from the mean
field equation,
\begin{align}
g_\sigma \left<
\bar{\psi} \psi\right>+ \Box \bar{\sigma} + U' (\bar{\sigma}) = 0
.
\label{eq: MFeq}
\end{align}
By substituting Eq.~(\ref{eq: divide}) into the relativistic
Lagrangian~(\ref{eq: Lag}), we get,
\begin{align}
{\mathcal{L}} \left(x\right)
=& \bar{\psi} \left(x\right) \left\{i \gamma_\mu \partial^\mu
- M^\ast \right\}\psi \left(x\right) \notag \\
& - g_\sigma \bar{\psi}
\left(x\right)  \psi \left(x\right) \sigma' \left(x\right)
+ \frac{1}{2}
\partial_\mu \bar{\sigma} (x)
\partial^\mu \bar{\sigma} (x) - U\left(\bar{\sigma}\right)
\notag \\
& + \partial_\mu \sigma' \left(x\right) \partial^\mu \bar{\sigma} \left(x\right)
- U' \left(\bar{\sigma}\right) \sigma' \left(x\right) \notag \\
& + \frac{1}{2} \partial_\mu \sigma' \left(x\right) \partial^\mu \sigma' \left(x\right)
- \frac{1}{2} U''\left(\bar{\sigma}\right) {\sigma'}^2 \left(x\right) +
 O ({\sigma'}^3)
\label{eq: efflag1}
.
\end{align}
$O({\sigma'}^{3})$ indicates the terms of more than ${\sigma'}^{3}$ orders.
We neglect the higher order terms, $O({\sigma'}^3)$, i.e., the harmonic approximation.
In the above equation $M^\ast$ is the effective mass of a nucleon,
\begin{align}
 M^\ast = M + g_\sigma \bar{\sigma}.
\end{align}
In Eq.~(\ref{eq: efflag1}), we introduce the new mass parameter $\mu$ for the $\sigma '$ field,
\begin{align}
\mu^2 = U'' (\bar{\sigma}) = m_\sigma^2 + 2 g_2 \overline{\sigma} + 3 g_3
\bar{\sigma}^2
\label{eq: musigma}
.
\end{align}
Using this, the propagator for $\sigma '$ meson field can be written,
\begin{align}
D (x-y) = \int \frac{d k^4}{(2 \pi)^4} 
\frac{e^{-i k(x-y)}}{k^2 - \mu^2 + i \varepsilon}.
\label{eq: propa}
\end{align}
It is noted that the mass term in the $\sigma '$ meson propagator
 is $\mu$, not $m_\sigma$.
Using this, we can write down $\sigma'$ field,
\begin{align}
\sigma' (x)  = \int d^4 y D(x-y) \left( g_\sigma \bar{\psi} (y) \psi (y) 
+ \Box \bar{\sigma} (y)
+ U' (\bar{\sigma}) \right). 
\end{align}
Substituting this expression into Lagrangian~(\ref{eq: efflag1}), we get,
\begin{align}
\int d^4x {\mathcal{L}}_{eff}  = & \int d^4 x \left\{
\bar{\psi} (x) \left( i \gamma_\mu \partial^\mu
- M - g_\sigma \bar{\sigma} (x) \right) \psi (x)
+ \frac{1}{2} \partial_\mu \bar{\sigma} (x) 
\partial^\mu \bar{\sigma} (x) - U (\bar{\sigma}) \right\}
\notag \\
&- \iint d^4 x d^4 y \left\{ g_\sigma \bar{\psi} (x) \psi (x) + \Box 
\bar{\sigma} (x) + U' (\bar{\sigma}) \right\}
\notag \\
& \times D(x-y)
\left\{ g_\sigma \bar{\psi}  (y) \psi (y) + \Box \bar{\sigma} (y) 
+ U' (\bar{\sigma}) \right\}
.
\end{align}
We may consider this effective Lagrangian as obtained by integrating out
the fluctuation field $\sigma'$.
For the interaction part we use the Gor'kov method \cite{KR91,FW71}, which is used 
in the theory of superconductor ,
\begin{align}
\int d^4x {\mathcal{L}}_{int} & = -  \iint d^4 x d^4 y D(x-y)
 \left\{ g_\sigma \bar{\psi} (x) \psi (x) + \Box \bar{\sigma} (x)
+ U' (\bar{\sigma} (x)) \right\} 
\notag \\
 & \times \left\{ g_\sigma \left<\bar{\psi} (y) \psi (y)\right>
+ \Box \bar{\sigma} (y) + U' (\bar{\sigma} (y)) \right\}
\notag \\
& - g_\sigma^2 \iint d^4 x d^4 y  D(x-y)
\bar{\psi} (x) \left<\psi (x) \bar{\psi}(y)\right> \psi (y)
\notag \\
& + \frac{1}{2} 
g_\sigma^2 \iint d^4 x d^4 y D(x-y)
\bar{\psi} (x) \bar{\psi} (y) \left<\psi (x) \psi(y)\right>
\notag \\
& + \frac{1}{2} g_\sigma^2 
\iint d^4 x d^4 y  D(x-y)
\left<\bar{\psi} (x) \bar{\psi} (y)\right> \psi (x) \psi (y)  .
\label{eq: Lint}
\end{align}
In the above equation, the first term  vanishes because of the mean field equation (\ref{eq:
MFeq}).
The second term corresponds to the Fock term and
the last two terms are responsible for the pairing interaction.
Finally, we get a modified effective Lagrangian of the form,
\begin{align}
 {{\mathcal{L}'}}_{eff} = {\mathcal{L}}_{MF} + {\mathcal{L}}_{int}
.
\end{align}
In this Lagrangian ${\mathcal{L}}_{MF}$ corresponds to the Lagrangian of
the RMF theory,
\begin{align}
{\mathcal{L}}_{MF} =   
\bar{\psi} (x) \left( i \gamma_\mu \partial^\mu - M - g_\sigma \bar{\sigma} (x)
\right) \psi (x) + \frac{1}{2} \partial_\mu \bar{\sigma} (x) 
\partial^\mu \bar{\sigma} (x) - U (\bar{\sigma}),
\end{align}
with which we can reproduce the RMF theory.
${\mathcal{L}}_{int}$ is given in eq.~(\ref{eq: Lint}).

Using the Euler-Lagrange equation to this Lagrangian, with respect to
$\psi$ and $\bar{\psi}$
we get the Dirac equations for the nucleon field $\psi$ and
$\bar{\psi}$,
\begin{subequations}
\begin{align}
(i \gamma_\mu \partial^\mu - M^\ast)_{ab} \psi_b (x) =& 
g_\sigma^2 \int d^4 y D(x-y) \left<\psi_a(x) \bar{\psi}_b (y)\right> \psi_b (y)
\notag\\
&- g_\sigma^2 \int d^4y D(x-y)
\left<\psi_a (x) \psi_b (y)\right> \bar{\psi}_b (y)
,
\end{align}
\begin{align}
(-i \gamma_\mu \partial^\mu - M^\ast)_{ba} \bar{\psi}_b (x) =& 
g_\sigma^2 \int d^4 y D(x-y) \left<\psi_b(y) \bar{\psi}_a (x)\right>
\bar{\psi}_b (y)
\notag \\
&- g_\sigma^2 \int d^4y D(x-y)
\left<\bar{\psi}_a (x) \bar{\psi}_b (y)\right> \psi_b (y).
\end{align}
\end{subequations}
With these equations we get the relativistic Hartree-Fock-Bogoliubov 
equation as written in Kucharek et al. 
\begin{align}
\begin{pmatrix}
h-\lambda &\Delta\\
- \Delta^\ast & - h^\ast + \lambda
\end{pmatrix}
\begin{pmatrix}
U_\nu \\ V_\nu
\end{pmatrix}
 = \varepsilon_\nu \begin{pmatrix} U_\nu \\ V_\nu \end{pmatrix}.
\label{eq: RHFB}
\end{align}
Here, $h$ is the Dirac Hamiltonian and $\Delta$ is
pairing field,
\begin{subequations}
\begin{align}
 h_{ab}\left(\boldsymbol{x},\boldsymbol{y}\right)
&= \left[-
i\boldsymbol{\alpha}_{ab}\cdot\boldsymbol{\nabla}_{\boldsymbol{x}}
- \gamma^0_{ab} M
+ \sum_\phi \int d^3 z \left(\gamma^0
 V^\phi\left(\boldsymbol{x},\boldsymbol{z}\right)
\right)_{acbd}
\rho_{dc}
\left(\boldsymbol{z},\boldsymbol{y}\right)
\right]\delta \left(\boldsymbol{x}-\boldsymbol{y}\right)
\notag \\
&\quad - 
\sum_\phi \int d^3 z \left(\gamma^0
V^\phi\left(\boldsymbol{x},\boldsymbol{z}\right)
\right)_{acdb} \rho_{dc}
\left(\boldsymbol{z},\boldsymbol{y}\right),\\
\Delta_{ab}\left(\boldsymbol{x},\boldsymbol{y}\right) & =
\sum_\phi \int d^3 z \left(\gamma^0
 V^\phi\left(\boldsymbol{x},\boldsymbol{z}\right)
\right)_{abcd}
\kappa_{cd}
\left(\boldsymbol{y},\boldsymbol{z}\right)
.
\end{align}
Here, we have included the chemical potential $\lambda$ which is fixed
by
$<\Tilde{A}|\psi^\dagger\psi|\Tilde{A}> = N$ with N being the particle number.
In the above equations,
$\rho$ and $\kappa$ are the normal density and the pairing density.
\end{subequations}
\begin{subequations}
\begin{align}
 \rho_{ab} \left(\boldsymbol{x},\boldsymbol{y}\right) &= \bigl<
\Tilde{A}\bigr|\psi_b^\dagger\left(\boldsymbol{y}\right)
 \psi_a\left(\boldsymbol{x}\right)\bigr|\Tilde{A}\bigr>,\\
 \kappa_{ab} \left(\boldsymbol{x},\boldsymbol{y}\right) &= \bigl<
 \Tilde{A}\bigr|\psi_b\left(\boldsymbol{y}\right)
 \psi_a\left(\boldsymbol{x}\right)\bigr|\Tilde{A}\bigr>
.
\end{align}
\end{subequations}
$V$'s are the meson potentials, which are introduced in the same
procedure as the case of the $\sigma$ meson:
\begin{align}
 V_{abcd}^\phi(\boldsymbol{x},\boldsymbol{y})
&= - g_\phi^2\left(\gamma^0\Gamma^\phi\right)_{ac}
 \left(\gamma^0\Gamma^\phi\right)_{bd}
\int \frac{d^3 k}{\left(2 \pi\right)^3}
 \frac{e^{-\boldsymbol{k}\cdot\left(\boldsymbol{x}-\boldsymbol{y}\right)}}
 {k^2 + \mu_\phi} \quad \left(\phi = \sigma, \omega, \rho, A\right)
.
\end{align}
$\Gamma^\phi$ is the nucleon-nucleon-meson vertex, for example,
$\Gamma^\omega = \gamma^\mu$.
The only difference between Kucharek et al.  and ours is that in our
formalism the masses of mesons in p-p
channels are modified by non-linear terms of mesons. (see Eq~(\ref{eq:
musigma}).)
\section{Result}\label{sec: result}
In this section we apply the formalism derived in the previous
section
to nuclear matter.
Before we mention the result, we explain the approximations that we use
in this work.
First, we neglect the Dirac sea as usually done in the RMF theory.
Second, we neglect the Fock term appeared in Eq.~(\ref{eq: Lint}) as
usual because
our aim is to check the effect of the non-linear terms of mesons on the
pairing channel.
Third, we take into account only the $^1S_0$ pairing, which is the most
important one in both nuclear matter and finite nuclei.
We do not treat proton-neutron pairing.
With these approximations, we deduce from Eq.~(\ref{eq: RHFB})
the BCS-type gap equation \cite{KR91},
\begin{align}
  {\Delta \left(k\right)} = - \frac{1}{8 \pi^2} \int_0^\infty
  {v_{pp}}\left(k,p\right)
  \frac{{\Delta \left(p\right)}}
{\sqrt{\left(\varepsilon \left(p\right)
  -\lambda\right)^2+{\Delta \left(p\right)^2}}} p^2 dp.
\label{eq: gapeq}
\end{align}
Here, $\Delta(k)$ is the pairing gap in the nuclear matter at momentum
$k$.
$\lambda$ is the chemical potential.
$v_{pp}(k,p)$ is the interaction term of the pairing channel.
As for the precise form of $v_{pp}(k,p)$, see Kucharek et al \cite{KR91}.
Furthermore, we introduce the form factor for each
vertex,
\begin{align}
 F_\alpha(q^2) = \frac{\Lambda_\alpha^2-m_\alpha^2}{\Lambda_\alpha^2-q^2} \quad \text{(
$\alpha$ = $\sigma$, $\omega$ and $\rho$)}
\end{align}
to cut out the effect of the high momentum region.
Without the cut-off, we get unreasonably large pairing
gap \cite{KR91,Ma98}.
In this work we use the cut-off parameter $\Lambda$, $1800$MeV for $\sigma$,
$1500$MeV for $\omega$ and  $1300$MeV for $\rho$.
These form factors correspond to those of Bonn C \cite{BM90}.
We note that there is large cut-off dependence of the value of the
pairing gap.
However, we find that the overall tendency is not changed.
The actual form of $v_{pp}$ is plotted in Fig~\ref{fig: vpp}.
In this figure, we also plot the components of the $\sigma$ and $\omega$
meson exchanges.
We use the TM1 parameter set \cite{ST94}.
From this figure, we see that the large attraction of $\sigma$ and
the large repulsion of $\omega$ cancel out each other and the relatively week
interaction in the lower momentum region is realized.
It is one of the characteristic feature of the RMF theory.

In Fig.~\ref{fig: delta_tm1} we show the pairing gap at the Fermi
momentum $k_F$ fm$^{-1}$ 
as a function of the density measured in the Fermi momentum.
We use the parameter set TM1, which has both $\sigma$
non-linear coupling and $\omega$ non-linear coupling terms \cite{ST94}.
In this figure, ``NL''(solid line) denotes the case when
we use the formalism in the
previous section to include the non-linear term effect on the pairing
channel and ``L''(dashed line) means the case with the use of the raw mass
$m_\sigma$, $m_\omega$ for the mass
parameters in the propagators of mesons instead of
$\mu_\sigma$ and $\mu_\omega$ (see Eq.~(\ref{eq: propa})).
From this figure we see that the pairing gap is reduced by a few
MeV by including the non-linear term of mesons.
This effect is especially large around $k_F$ = 1.0 fm$^{-1}$.
Only the difference between two cases is the mass terms in the
propagator of mesons.
In our formalism,
by including the non-linear term effect the masses of mesons in the
propagator are modified with the medium effect. 
To see it, we plot $\mu_\sigma$ (solid line) and $\mu_\omega$ (dashed
line)
in Fig.~\ref{fig: mass_tm1}.
At lower density both $\mu_\sigma$ and $\mu_\omega$ are not much
different  from
their original values, $m_\sigma$ and $m_\omega$,
because the densities are low,  medium effect is not large in this region.
However, at higher density, they are larger than their original values.
These results come from the non-linear terms in the expressions of
$\mu_\sigma$ and $\mu_\omega$ (Eq.~(\ref{eq: musigma})).
Those effects are more significant at higher density.
Because the mass terms in the propagator are more massive at higher
density region, the interaction in the pairing channel is reduced.
To see that in Fig.~\ref{fig: vpp} we compare the value of $v_{pp}$
at the density $k_F$ = 0.78 fm$^{-1}$, which corresponds to the maximum
point of $\Delta(k_F)$ without the non-linear term effect in
Fig.~\ref{fig: delta_tm1}.
We see that by introducing the non-linear term effect, $v_{pp}$ is
reduced.
Hence, pairing gaps are reduced.
Of course, those effects depend on the value of coupling constant.
However, at normal density the general forms of non-linear terms are not 
so different between the parameter sets which are widely used.
So, we can find the same tendency; namely, the reduction of pairing gap with
each parameter set at moderate density.

We study the effect of the form factor by changing the cut-off mass slightly.
With the use of $\Lambda_\sigma=$2200MeV, $\Lambda_\omega=$1900MeV,
$\Lambda_\rho$=1700MeV, the pairing gap increases largely as
shown in Fig~\ref{fig: delta_check}.
The peak position, however,  does not change much from the previous
choice of the
parameters. 
These cut-off masses in the corresponding form factors are related with
the structure of hadrons in the nuclear medium. It is certainly an
interesting future subject to fix these values on theoretical ground.
 
For comparison we show $\Delta(k_F)$ for the case of the NL1 \cite{RRM86}
parameter set in Fig.~\ref{fig: delta_nl1}.
NL1 has the non-linear term of only $\sigma$ meson. It does not have those
of $\omega$ meson.
We can see that the same effect appears as in the case with TM1.
However, at higher density $\Delta(k_F)$ get abnormally large values.
(It is not shown in the figure.)
This is due to the minus sign of the coupling constant of the quadratic
self-coupling term.
Because of this negative sign, at higher density $\mu_\sigma$ is getting 
smaller and smaller with density.
Hence, the pairing interaction becomes very large and
the pairing gap becomes very large at high densities.
In the TM1 case, because $g_3$ has positive sign this effect does not occur.
Nevertheless, we can find the same tendency in the moderate mass region.
\section{Conclusion}\label{sec: conc}
We have developed a new framework to incorporate non-linear terms 
of mesons in pairing channel in the RHFB theory.
We have applied the RHB theory to 
nuclear matter to study the effect of non-linear terms
on the p-p channel.
We have found that the non-linear terms reduce $\Delta$ at Fermi momentum
especially for normal density region.
It is caused by non-linear term effect on the mass parameter in the meson
propagator.
The pairing interaction is reduced due to the non-linear coupling terms.
This effect is neglected in other works.
We have found that it has a non-negligible effect.
It is interesting to apply it to asymmetric nuclear matter and finite 
nuclei.

In this paper we have used the parameter sets determined in the RMF
framework; i.e., TM1 and NL1. As seen in the formalism, the pairing
correlations in nuclear matter are caused by the fluctuations of the
meson fields. Hence, it is very natural to take the Fock term when we
consider the pairing correlations. It then needs an elaborate work to
complete our program as to first fix the parameters of the relativistic
Lagrangian in the Hartree-Fock framework from various magic number
nuclei and then to calculate the pairing properties of nuclear
matter. We start this program to be worked out in due time.
%

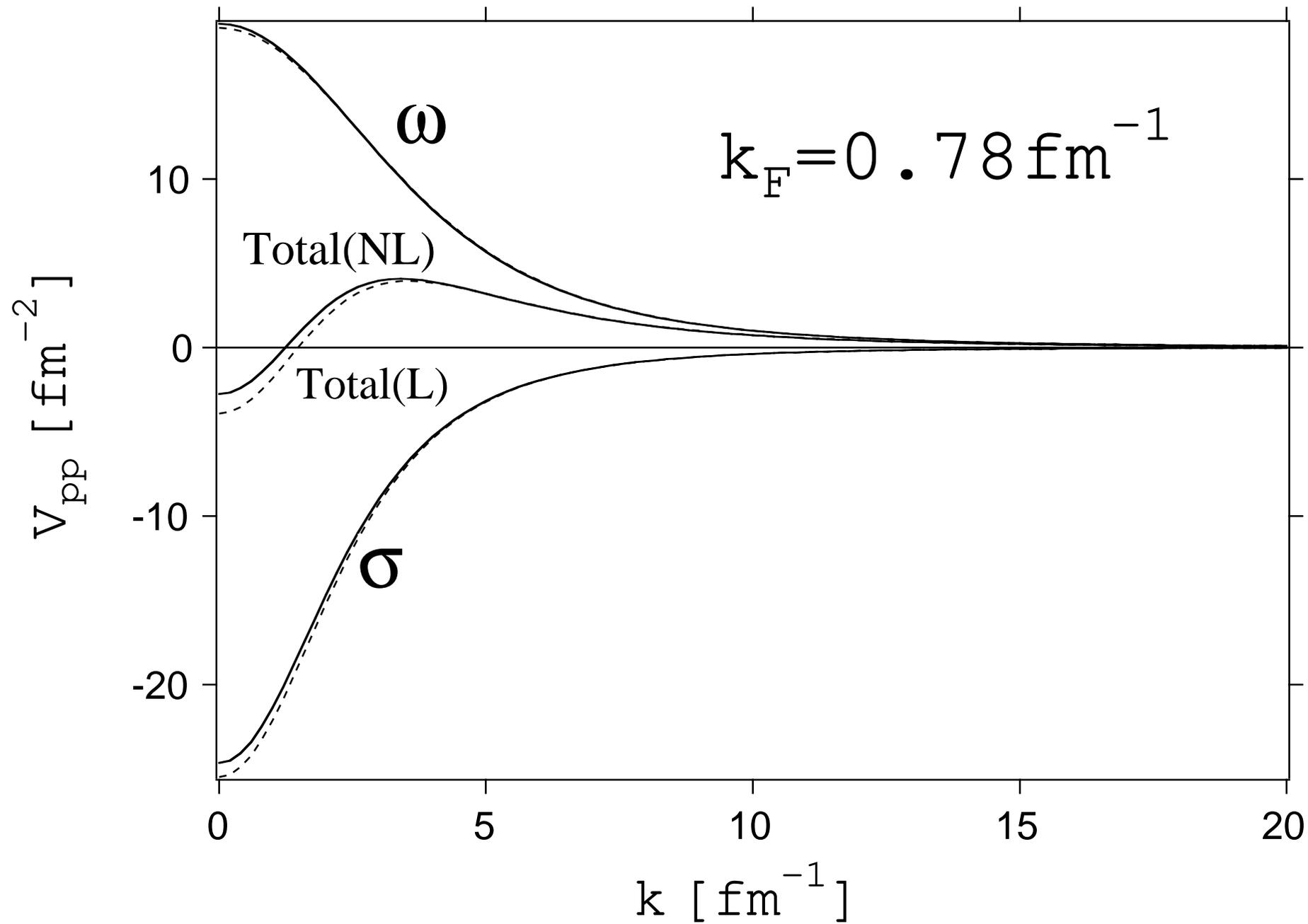
\begin{figure}
\caption{
The interaction kernel in the gap equation $v_{pp}(p,k)$ at p 
= 0.78 fm$^{-1}$ for the case of the Fermi momentum, $k_F$ = 0.78
 fm$^{-1}$.
The horizontal axis is $k$ in fm$^{-1}$.
``NL''(solid line) denotes the case of non-linear
terms of mesons being incorporated and ``L''(dashed line) denote the
 case without them.
``Total'' corresponds to  the sum of the contributions from $\sigma$,
$\omega$, and $\rho$ mesons, where the curves with $\sigma$ and $\omega$
 denote the contributions
from $\sigma$ and $\omega$ mesons, respectively.
\label{fig: vpp}
}
\end{figure}
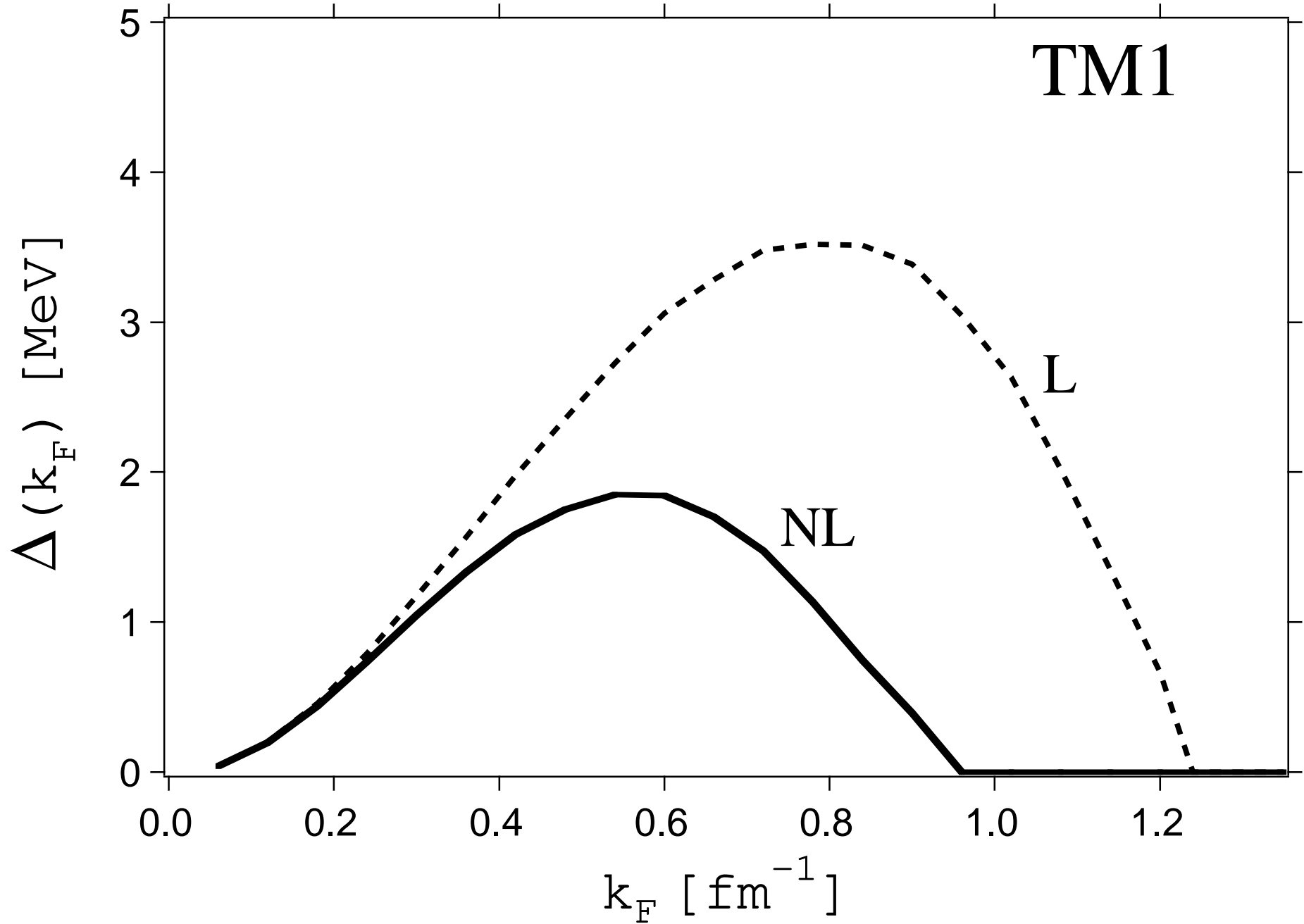
\begin{figure}
\caption{
Pairing gap $\Delta$ at the Fermi momentum $k_F$ as a
function of the Fermi momentum.
The TM1 parameter set is used.
``NL''(solid line) corresponds to the case in which the
non-linear term of mesons are implemented.
``L''(dashed line) corresponds to the case without them.
}
\label{fig: delta_tm1}
\end{figure}
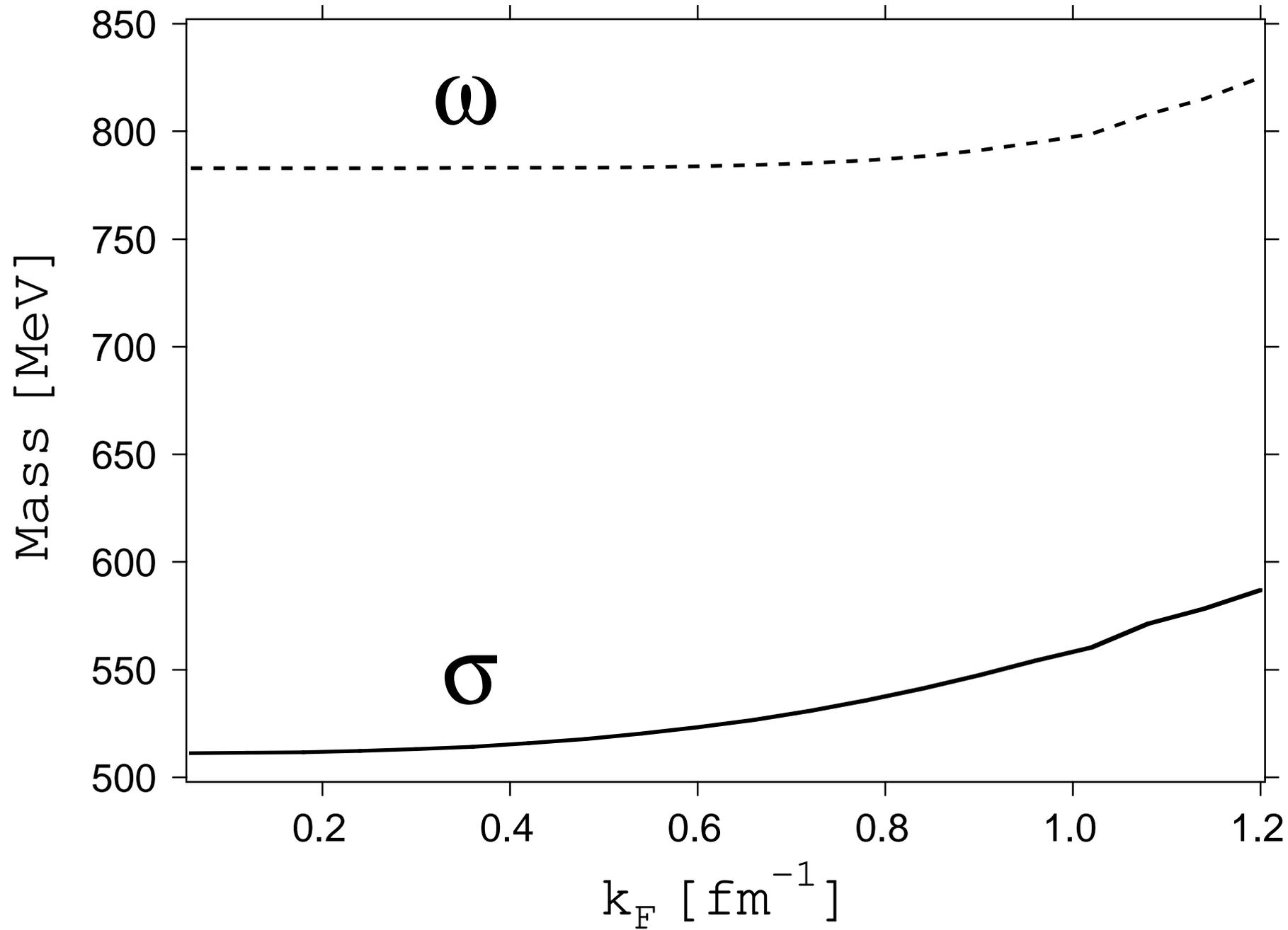
\begin{figure}
 \caption{
The mass parameters of $\sigma$ and $\omega$ mesons in the
propagators, $\mu_\sigma$ (solid line) and $\mu_\omega$ (dashed line).
The horizontal axis is the Fermi momentum $k_F$ fm$^{-1}$.
The increases of masses due to the non-linear terms of mesons are
 observed at higher Fermi momentum.\\
}
\label{fig: mass_tm1}
\end{figure}
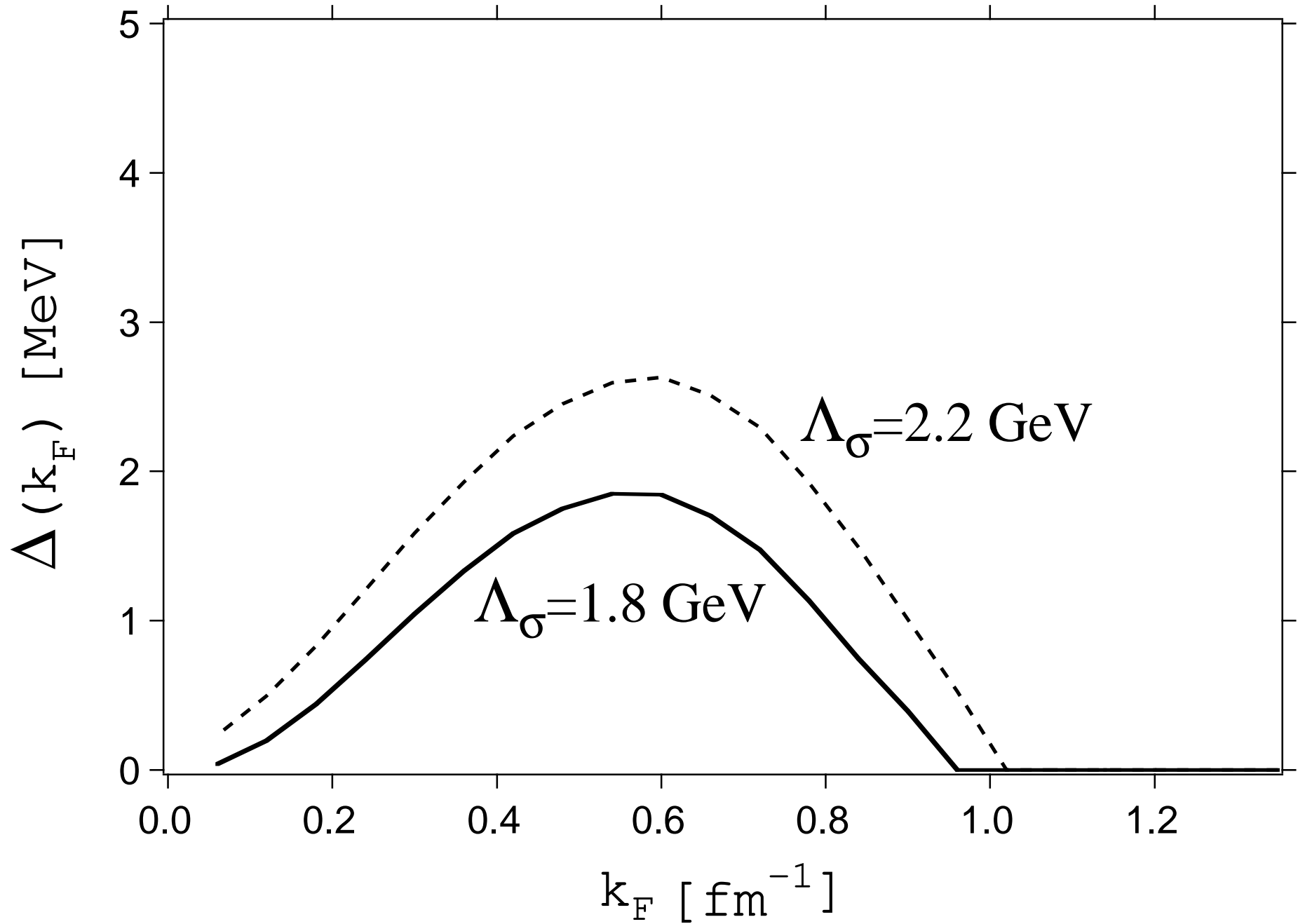
\begin{figure}
 \caption{
Pairing gap $\Delta$ at the Fermi momentum $k_F$ as a
function of Fermi momentum.
The TM1 parameter set is used.
Dashed line corresponds to the cut-off parameters, $\Lambda_\sigma$=2200MeV,
$\Lambda_\omega$=1900MeV, $\Lambda_\rho$=1700MeV.
Solid one corresponds to the cut-off parameters, $\Lambda_\sigma$=1800MeV,
$\Lambda_\omega$=1500MeV, $\Lambda_\rho$=1300MeV.
}
\label{fig: delta_check}
\end{figure}
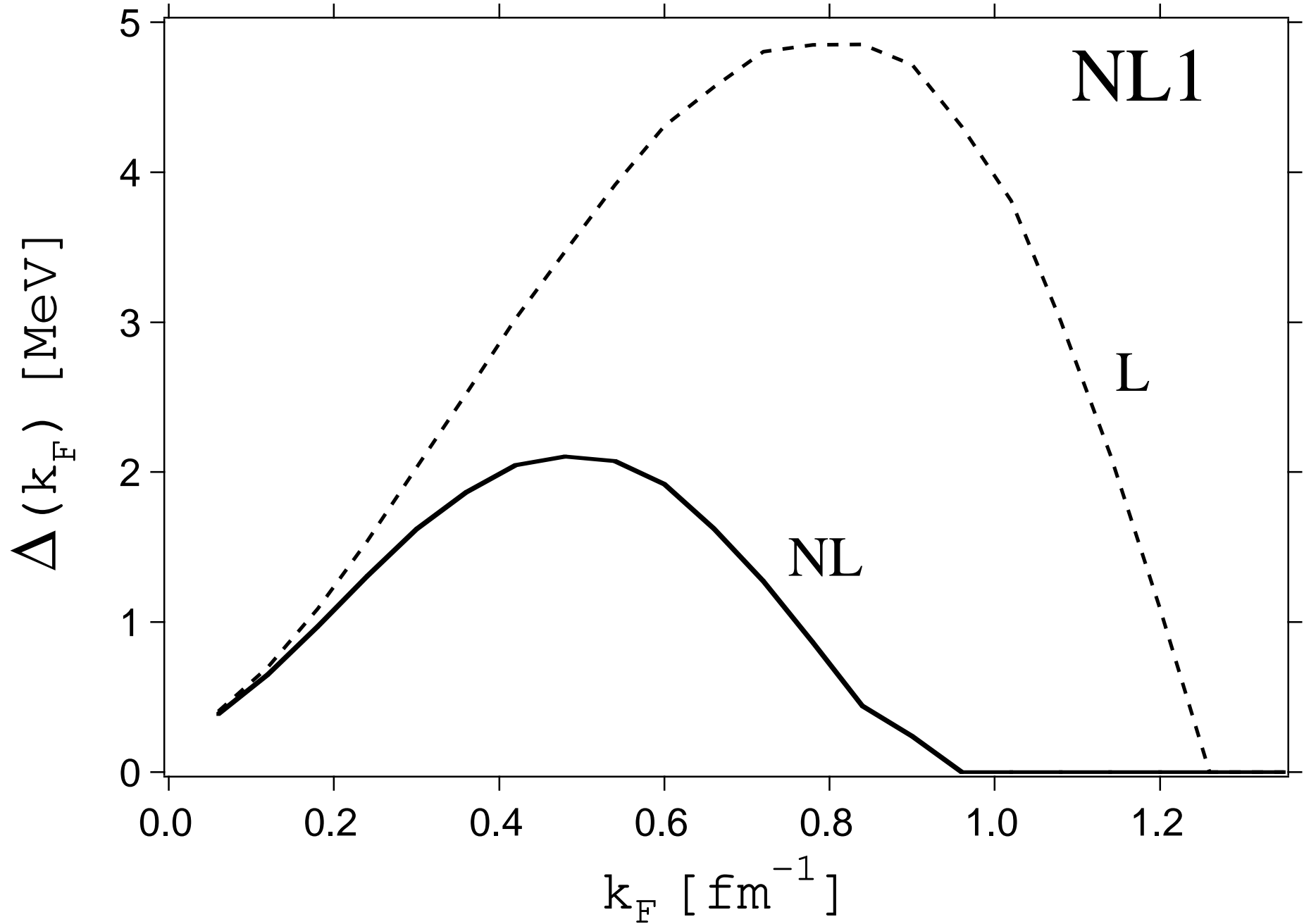
\begin{figure}
 \caption{
Pairing gaps $\Delta$ at the Fermi momentum $k_F$ as a
function of Fermi momentum for the case of the NL1 parameter set.
The solid curve corresponds to the case with non-linear term, whereas
 the dashed curve without the non-linear term.
}
\label{fig: delta_nl1}
\end{figure}
\end{document}